\input harvmac

\Title{hep-ph/9608197, SCIPP 96/31, RU - 96 /95}
{\vbox{\centerline{The Cosmology of String Theoretic Axions}
}}
\bigskip
\centerline{Tom Banks$^a$ and Michael Dine$^b$}
\smallskip
\centerline{\it $^a$Department of Physics, Rutgers
University, Piscataway, NJ 08540 }
\centerline{\it $^b$Santa Cruz Institute for Particle Physics,
 University of California, Santa Cruz, CA 95064}

\bigskip
\baselineskip 18pt

\noindent
String theory posesses numerous axion candidates.
The recent realization that the compactification radius
in string theory might be large means that these  states
can solve the strong CP problem.   This still leaves
the question of the cosmological bound on the axion mass.
Here we explore two schemes for accommodating such light
axions in cosmology.  In the first, we note that in string
theory the universe is likely to be dominated early
on by the coherent oscillations of some moduli.
The usual moduli problem  assumes that these fields
have masses comparable to the gravitino.
We argue that
string moduli are likely to be substantially more massive,
eliminating this problem.
In such cosmologies the axion bound is significantly
weakened.  Plausible mechanisms for generating
the baryon number density are described.
In the second, we point out that in string theory,
the axion potential might be much larger at early
times than at present.  In string theory,  if CP violation
is described by a small parameter,
the axion may sit sufficiently close to its true minimum
at early times
to invalidate the bounds.

\Date{7/96}

\newsec{Introduction}

The invisible axion is an elegant solution to the strong CP problem.  At first glance, 
string theory is replete with axion candidates.  At weak coupling,
these include the ``model-independent axion"\ref\wittenaxion{
E. Witten, Phys. Lett. {\bf 149B} (1984) 351.} and axions which
arise from internal components of
$B_{\mu \nu}$\ref\newissues{E. Witten,
Nucl. Phys. {\bf B268} (1986) 79.}.
There are two problems with these axions.
First, in the weakly coupled region there are no
good arguments that QCD is the dominant contribution
to the potential of any
axion.  Consider, first, the usual ``model-independent"
axion.  While this axion respects a Peccei-Quinn (PQ)
symmetry in perturbation theory, this symmetry
is likely to be broken by both stringy non-perturbative
effects and hidden sector dynamics.  In ref.
\ref\coping{T. Banks and M. Dine, ``Coping with
Strongly Coupled String Theory,"
Phys. Rev. {\bf D50} (1994) 7454,
hep-th/9406132.}, it was shown that discrete symmetries
could adequately suppress hidden sector contributions.
However, we know of no argument that there cannot
be inherently stringy PQ
symmetry-violating effects of order $e^{-c/g}$.  Unless
$c$ was suprisingly large, these effects would be far larger
than those associated with QCD.
Similarly, at large radius,
there are potential axions associated
with the internal components of the antisymmetric tensor.
These symmetries, however, are broken by world-sheet
instantons, the breaking being of order $e^{-R^2}$.  In the weak
coupling picture, however, $R$ is necessarily of order $1$.

The second problem with these axions is that their
decay constants appear to be incompatible
with cosmological bounds\ref\axioncosmologya{J. Preskill,
F. Wilczek and M. Wise, Phys. Lett. {\bf 120B} (1983) 127.}
\ref\axioncosmologyb{L. Abbott and
P. Sikivie, Phys. Lett. {\bf 120B} (1983) 133.}\ref\axioncosmologyc{M. Dine and W. Fischler,
Phys. Lett. {\bf 120B}(1983) 137.}.
We have recently shown that
in certain strongly coupled string vacua, the first difficulty is
removed\ref\mtheory{T. Banks and M. Dine,
``Couplings and Scales
in Strongly Coupled Heterotic String
Theory," RU-96-27, hep-th/9605136.}.  The purpose
of the present paper is to demonstrate that the cosmological argument is not a barrier to
stringy axion models.\foot{There have been previous proposals to solve
this problem.  Late decaying particles, somewhat similar to the moduli under
discussion here, have been considered in ref. \axioncosmologyc\
and in \ref\turneretal{P.J. Steinhardt and
M.S. Turner,Phys. Lett. {\bf 129B} (1983) 51.} and
\ref\kmy{M. Kawasaki,
T. Moroi and T. Yanagida, ``Can Decaying
Particles Raise the Upperbound on the
Peccei-Quinn Scale?", hep-ph/9510461.} .
Weak anthropic ideas have been considered
in \ref\linde{A.D. Linde,
Phys.Lett. {\bf 201B} (1988) 437.}.}
In fact, we will present two different scenarios
for cosmology and particle physics in which
axions with decay constants much larger than the conventional bounds are allowed.
The first scenario depends on a number of relatively obscure facts about string theory.
It is conventional to estimate the order of magnitude of the decay constant of stringy
axions, as well as analogous parameters for other moduli, to be the Planck scale.  
In fact, in a calculation done almost ten years ago, Kim and Choi
\ref\choi{K. Choi
and J.E. Kim,
Phys. Lett. {\bf 154B} (1985) 393; K. Choi and
J.E. Kim, Phys. Lett. {\bf 165B} (1985) 71.} showed that the model
independent axion decay constant in weakly coupled heterotic string theory is really
${M_p \over 16\pi^2}$ or about $10^{16}$ GeV.  So it is not unreasonable to expect large
numerical factors in the relation between decay constants and the Planck mass.  
In the strong coupling regime, we have suggested an effect which might further lower
the decay constant.   

It does not seem plausible, however,
that such arguments can lower the axion decay constant by
 the seven orders of magnitude necessary to satisfy the conventional
cosmological bound of $10^{12} \rm GeV$.
However, it is likely that string cosmology is not entirely conventional.  Indeed,in all 
vacuum states explored to date, there are moduli whose masses are determined by
supersymmetry (SUSY) breaking.  It is usually assumed that
the characteristic scales of variation of these moduli -- what we might
loosely refer to as their ``decay constants", are of order
the Planck mass.  In that case, they dominate the
energy density of the universe until its energy density is
too low for conventional 
nucleosynthesis to proceed.  However, as we will
see, a mild
retuning of the modular decay constants, similar
to,  but less dramatic than, that of Kim and Choi for
the axion, is sufficient to solve the moduli problem.
Typically this leads to a cosmology which
is matter dominated from a short time after 
inflation until just before nucleosynthesis.    

In such a context, we will see that the axion problem is greatly ameliorated.
We will make the standard assumption that after
inflation the axion is left stranded at a 
random place on its potential.  We then show that 
the axion begins to oscillate at a time when its energy 
density is smaller than the total density of the universe by a factor of 
${f_a^2 \over m_p^2}$.  If the cosmic energy density is dominated by other coherent scalars
(or nonrelativistic matter), this ratio remains constant until those scalars decay
into relativistic matter.  There is a wide range of reheat temperatures, above the 
nucleosynthesis temperature, for which axions with decay constants much larger than the
conventional bounds do not dominate the universe before the
conventional era of matter 
domination.  Furthermore, the scenario is compatible with the idea that the coherent
scalars whose decay gives rise to the Hot Big Bang are moduli fields,
that the
axions are the dark matter, and that gravitinos and topological
objects such as domain walls are sufficiently diluted.

The question that remains is the origin of baryogenesis.  We argue that the decay of the
moduli may well be the source of the baryon asymmetry.  Such a scenario requires dimension
four baryon number violating operators.  It is not consistent with the existence of a stable
superpartner, which might be the dark matter.  On the other hand, for a range of decay
constants axions can be the dark matter and a stable sparticle is unnecessary.  We also
study a scenario in which baryogenesis arises from coherent condensates of standard model
sparticle fields\ref\affdine{I Affleck and
M. Dine, Nucl. Phys. {\bf
B249} (1985) 361.}
\ref\drt{ M. Dine,
L. Randall and S. Thomas, ``Baryogenesis
{}from Flat Directions of the
Supersymmetric Standard Model," Nucl.
Phys. {\bf B458} (1996) 291, hep-ph/9507453.}.  We find that this possibility is
viable if the associated directions are lifted only
by very high dimension operators -- or not at all.

Our second strategy for relaxing the bound on the axion decay constant depends on an
unorthodox
assumption about the origin of CP violation.  The standard model of CP violation
through Cabibbo Kobayashi Maskawa (CKM) phases requires a fundamental CP violating phase of
order one.  The miniscule size of CP violating effects in current experiments is attributed
to the necessity of mixing all three quark generations, and to the small size of the CP 
conserving mixing angles.  In such a scheme one expects generic CP violating 
phenomena to be unsuppressed.  In the context of supersymmetry,
one has difficulty understanding the smallness of CP-violating
phases in soft breaking parameters required by phenomenology,
and the smallness of $\theta$ is a mystery.   
There exist alternate approaches in which the fundamental CP
violating phases are all small, so that the breaking of CP is everywhere controlled by a small
parameter.  We will review such a proposal, due to Nir and
Ratazzi \ref\nir{Y. Nir and R. Rattazzi,
``Solving the Supersymmetric CP Problem
with Abelian Horizontal Symmetries,"
RU-96-11, hepph/9603233.}.  In this class of models, gaugino
and other phases are automatically small enough, a CKM phase of
conventional size explains all current data on CP violation,
but
the smallness of $\theta$ requires further explanation.

String theory is a theory where $CP$ is a good symmetry --
in fact a gauge symmetry -- which must be spontaneously
broken\ref\stringcp{K.-W. Choi,
D.V. Kaplan and A.E. Nelson,
Nucl. Phys.
{\bf B391} (1993) 515; M. Dine,
R.G. Leigh and D. MacIntire,
Phys. Rev. Lett. {\bf 69} (1992) 2030.}.
It contains a host of CP odd fields which might provide the
small breaking of \nir.  Moreover it contains numerous axion
candidates which can relax $\theta$ to zero. The usual
objections to such a picture would be the large value
of the axion decay constants, and domain walls due to spontaneous CP violation.
Since CP is a gauge symmetry, the domain walls are not absolutely
stable, but they may have lifetimes much longer than the age of the universe.
In the conclusions we will argue that this and all other domain wall problems
can be solved in models where the energy density is matter dominated until quite
low scales.  In such models, spontaneous symmetry breaking is frozen in at
the end of inflation and the temperature never gets high enough for symmetry
restoring phase transitions.

Moreover,
in a world with everywhere small, spontaneous
CP violation, the cosmological axion problem may be
significantly ameliorated.
During inflation, certain {\it inflaton} fields remain
displaced from their minima.  As a consequence, the effective potential for all other
scalars may take values significantly different than their values in the vacuum.  
These, {\it passive}, scalars are rapidly driven to the minima of the inflationary
potential.  In general, the inflationary minimum will be nowhere near the true minimum 
of the vacuum potential.  (This is, in some sense, the
origin of the ``moduli problem.")  Among these
passive fields are the candidate axions.  Their
potentials can be large either if the
QCD scale is large due to a displaced dilaton\ref\dvali{G. Dvali,
``Removing the Cosmological Bound on the Axion Scale," 
IFUP-TH-21-95, hep-ph/9505253.},\foot{The proposal of
\dvali\ suffers from a number of difficulties.  These have
been discussed recently in \ref\choidvali{K. Choi, H.B. Kim and J.E. Kim,
``Axion Cosmology with a Stronger QCD in the
Early Universe,'' hep-ph/9696372.}.
Our proposal is quite different
in that, first, we explain why the early minimum
coincides with the minimum at late times, and
the QCD scale is assumed much larger, so that there
is little or no suppression of the axion mass.} 
or if world sheet instanton effects (in weakly coupled string
language) are enhanced due to diplaced Kahler moduli.
However, if the inflaton fields
are all CP invariant, and
CP is broken only by small effects, then the minimum of the inflationary potential for CP-odd
fields like the axion will be close to its true minimum.\foot{We will neglect
the possibility that the inflationary minimum is at $a = \pi$ while the true minimum
is $a = 0$.  It is clearly possible to construct models in which this alternative is
realized.}  Thus, during inflation, the axion is driven very close to its true minimum by
a very large potential.  It is easy to see that the postinflationary axion energy density
will be of order $\delta^2 \Lambda_{QCD}^4$, where $\delta$ is the small parameter
which controls CP violation.  We will see that for values of $\delta$ compatible with
experiment, this is small enough to significantly enlarge the allowed range of values
for the axion decay constant.

In the next two sections of this paper we present the details of the two scenarios
outlined above.  A third section is devoted to a brief discussion of models based on
eleven dimensional supergravity which sparked our reexamination of the axion bound.
The final section is devoted to our conclusions.

\newsec{Axions in Moduli Dominated Cosmologies}

\subsec{Moduli Ameliorate the Axion Problem}

In our first scenario, we imagine that the postinflationary universe is dominated by one or
more moduli fields, with reheat temperature $T_R$.  From the end of inflation, until
energy densities of order $T_R^4$ the universe is matter dominated.  The axion begins this
era at a position determined by its inflationary potential, generically a distance of
order one (we define the axion to be dimensionless and to
have period $2\pi$) from its
true minimum.  Initially, it contributes a negligible amount, of order $\Lambda_{QCD}^4$ 
to the cosmic energy density.  The axion remains more or less stationary until the Hubble
parameter $H$ is equal to the axion mass.  After this time, it behaves like nonrelativistic
matter.  The crossover occurs when
\eqn\cross{{\sqrt{\rho} \over m_p} = {\Lambda_{QCD}^2 \over f_a}}
Notice that it is the reduced Planck mass $m_p = 2 \times 10^{18}$
GeV which enters 
this equation.  

At this time, the ratio of axion to modular energy densities is of order
\eqn\ratio{{\rho_a \over \rho} \approx {\Lambda_{QCD}^4 \over \rho} \sim {f_a^2 \over m_p^2}}
The ratio retains this value until reheat, at which time the axion energy density 
begins to grow relative to the energy of the massless particle gas produced by the decay
of the modulus.  The growth is linear in the inverse temperature.  

Classical cosmology begins with the era of nucleosynthesis, where the photon temperature 
is of order $1$ MeV.  In order to preserve the physics of this era, we
must have $T_R > 1$ MeV.\foot{Actually, one needs a somewhat
larger reheat temperature, of order $6$ MeV\ref\sixmev{M. Reno and D. Seckel, Phys.
Rev. {\bf D37} (1988) 3441; G. Lazarides,
R. Schaefer, D. Seckel and Q. Shafi,
Nucl. Phys. {\bf B346} (1990) 193.}.}
We must also arrange the correct baryon to entropy ratio, of
which more below.  Finally, we must ensure that the universe remains radiation
dominated until the conventional beginning of the matter dominated era
at $T \sim 10$ eV.
Since the axion to radiation ratio increases monotonically during this era, the only
requirement is that the ratio must be less than or equal to one when the radiation 
temperature is $10$ eV. Thus
\eqn\bound{{f_a^2 \over m_p^2}{T_R \over {10\ eV}} \leq 1}
Equality corresponds to the interesting case in which axions are the dark matter in
the universe.  

The borderline case $T_R \sim 1$ MeV gives an axion decay constant bounded 
by $6 \times 10^{15}$
GeV.  Given the results of Kim and Choi \choi , this seems to be right in the
interesting range for superstring axions.  The bound decreases like
$T_R^{-\ha}$ as $T_R$ 
increases.  In particular, if we take $T_R \geq 100$ GeV ,
as would be required for
baryogenesis at the electroweak phase transition, then $f_a \leq 10^{13}$ GeV.  While
this is still higher than the conventional bound, it seems a rather low number to expect to
come out of string theory.  Perhaps in the strong coupling vacuum described
in\ref\horava{P. Horava and
E. Witten, ``Heterotic and Type I String Dynamics From
Eleven Dimensions'', {\it Nucl. Phys.} {\bf B460}, (1996)506, hep-th/9510209;
``Eleven-Dimensional Supergravity on a Manifold with
Boundary," hep-th/9603142.}\mtheory,
if the QCD axion is a {\it boundary modulus} we could obtain a decay constant this
small.  Dimensional analysis suggests $10^{16}$ GeV for the decay constant of such a
boundary axion, and numerical factors such as those of \choi\ could bring us the rest of
the way.  We have had to choose parameters 
at their extreme ranges of plausibility to make this scenario work.
Another mechanism of low energy baryogenesis seems to be indicated.

Before enquiring what that mechanism could be, let us indicate the expectations for $T_R$.
The modulus whose decay initiates the Hot Big Bang \foot{There may be several fields
which deserve this designation.  We will pretend that there is only one for linguistic
convenience, but nothing that we say depends on this assumption.}
has a potential of the form
$M^4 {\cal V}({\phi \over f})$ where ${\cal V}$ is a bounded function.  Here $M$ is 
\lq\lq the fundamental scale of string theory\rq\rq .  Recent developments in string duality
have shown us that we know much less about the value of $M$ than we previously
thought.  In weakly coupled heterotic string theory, the best fit to the real world has
$M$ of order the reduced Planck mass.  In the strong coupling regime\mtheory, $M$ 
is closer to the unification scale.  Furthermore, the function ${\cal V}$ can vary quite
a bit over moduli space.   What is relevant to our present discussion is the value of $
{\cal V}$ near the minimum of the lightest moduli (those with the lowest reheat temperature).
Let us factor this value out and include it in $M$.  Then, in the regime of interest,
${\cal V}$ is of order one or less.   It is plausible that for the lightest unstable
moduli the effective value of $M$ is the fundamental SUSY
breaking scale : $M \sim 
\sqrt{F} \sim 10^{11}$ GeV (corresponding to
$m_{3/2}$ of order a few hundred GeV).  This appears to be
the case in all explicit models of
which we are aware.  Then the mass of the modulus is of
order
\eqn\modulusmass{m^2 \approx {M^4 \over f^2}.} 

As a model for the decay of the moduli, consider the usual
dilaton, $D$.  This field (canonically normalized)
couples to photons through a term
\eqn\dilatonphoton{{\cal L}_{D\gamma}\approx{1 \over m_p} D F_{\mu \nu}^2.}
So the decay width is of order
\eqn\decay{\Gamma \approx {m^3 \over m_p^2}\approx {M^6 \over f^3 m_p^2}}
which gives a reheat temperature
\eqn\reheat{T_R \approx \sqrt{\Gamma m_p}
\approx  {M^3 \over m_p^{1/2}f^{3/2}}}
This is of order the electroweak transition temperature for $f \sim 10^{14}$
GeV.  
So, even given the results of \choi ,
an electroweak reheat temperature seems difficult to achieve.
On the other hand,  if
$f \sim 10^{16} GeV$,
the reheat temperature is of order $1$ GeV,
well above the nucleosynthesis bound.

It may seem troubling to contemplate such
small values of $f$.  These correspond to
moduli masses of order $10$-$100$ TeV,
several orders of magnitude larger than the gravitino mass.
One might expect that this requires fine tuning.
But, as we have argued elsewhere, such large
masses are almost inevitable\ref\coping{T. Banks and M. Dine,
``Coping with Strongly Coupled String Theory,''
Phys. Rev. {\bf D50} (1994) 7454, hep-th/9406132 }.  In particular,
for small values of the gauge coupling, the
superpotential typically behaves as
\eqn\typicalw{W\approx e^{-16 \pi^2 S/N},}
where $S$ is the dilaton supermultiplet and
$N$ is of order $4$ or $5$.  We have normalized
the dilaton multiplet here as in \ref\wittenreduction{E.
Witten, Phys. Lett. {\bf 155B} (1985) 151. }, so that the
dilaton has a canonical kinetic term at weak
coupling.  With this
normalization, $16 \pi^2 S$ is periodic
with period $2 \pi$. The resulting potential has no
minimum at weak coupling, and
one must assume that there are large corrections to the
Kahler potential in order to give a stable
vacuum.  The second derivative of the potential
is then, indeed, quite large.  Moreover, the large
corrections to the Kahler potential introduce significant
uncertainties in the decay rates.  We have allowed
for these uncertainties in our estimates above by neglecting
factors of $4\pi$, etc., which are usually included in weak-coupling
based analyses, so our formulas are somewhat more optimistic
than others which appear in the
literature\ref\randallthomas{L. Randall and S. Thomas,
``Solving the Cosmological Moduli Problem with
Weak Scale Inflation,''
Nucl. Phys.
{\bf B449} (1995) 229, hep-ph/9407248.}.

In view of these remarks,
it may be correct to conclude that the cosmological moduli
problem \ref\cosmoduli{T. Banks, D.B. Kaplan
and A.E. Nelson,
``Cosmological Implications of Dynamical Supersymmetry
Breaking," Phys. Rev. {\bf D49} (1994) 779.B.
de Carlos, J.A. Casas, F. Quevedo and E. Roulet,
Phys. Lett. {\bf B318} (1993) 447.}
is a red herring, resulting from overreliance on naive 
dimensional analysis.  While moduli will certainly modify cosmic history above
the electroweak phase transition, and quite possibly between this scale and nucleosynthesis,
there is no longer a strong reason to believe that they interfere with classical 
cosmology.

\subsec{Baryogenesis}

However, moduli will certainly modify baryogenesis, the gravitino
problem, and the
axion bound.  Indeed, it seems difficult to push the reheat temperature of moduli
whose masses come from SUSY breaking above the electroweak scale.  Thus, even if
electroweak baryogenesis remains a viable option in the presence of moduli, its details
will probably be modified by modular decay.  This appears to be a complicated problem, and
we will not explore it.  Rather, following our conclusions about axions, we will
now explore other possibilities for baryogenesis.

The most straightforward scenario is to assume that modular decay itself is responsible
for the baryon asymmetry.  This sort of mechanism for baryogenesis apparently
originates with the work of \ref\ramond{G.D.
Coughlan, G.G. Ross, R. Holman, P. Ramond,
M. Ruiz-Altaba and J.W.F.
Valle, Phys. Lett. {\bf 160B} (1985) 249; G.G. Ross, R. Holman, P. Ramond, {\it Phys. Lett.} {\bf 137B}
, 343, (1984).}.  Many of the important issues are reviewed
in \ref\thomasbaryons{S. Thomas, ``Baryons and Dark Matter from the
Late Decay of a Supersymmetric Condensate,'' Phys. Lett. {\bf B356} (1995)
256, hep-ph/9506274.}  
The couplings by which the modulus decays 
may contain CP violation and baryon number violation of relative order one \foot{Here
we {\it do not} make the assumption of small CP violation which will dominate our
discussion in the next section.}.  In order to produce a baryon asymmetry, we must have
another sort of baryon violating operator in the lagrangian 
\ref\weinberg{D.V.Nanopoulos, S.Weinberg, {\it Phys. Rev.} {\bf D20}, 2484, (1979).}.  We will
see that the coefficient of this operator cannot be too small, so it is natural to
assume that it is one of the allowed renormalizable baryon number violating operators
in the supersymmetric standard model.  It is well known that the presence of such
operators is compatible with the stability of the proton and the experimental absence
of neutron-antineutron oscillations \ref\raby{C.S.
Aulakh and R.N. Mohapatra, Phys.
Lett. {\bf 119B} (1982) 316;
L.J. Hall and M. Suzuki, Nucl. Phys.
{\bf B231} (1984) 419;
F. Zwirner, Phys. Lett. {\bf 132B} (1983) 103;
S. Dawson, Nucl. Phys. {\bf B261} (1985)
297; R. Barbieri and A. Masiero, Nucl.
Phys. {\bf B267} (1986) 679;
S. Dimopoulos and L.J. Hall, Phys. Lett.
{\bf B207} (1987) 210; L.
Hall, Mod. Phys. Lett. {\bf A5} (1990) 467;
K.S. Babu and R.N. Mohapatra, Phys. Rev.
{\bf D42} (1990) 3778.}.  It is not compatible with a model
of the dark matter as a stable supersymmetric particle.  However, in
the present context, axions can play the role of dark matter,
and there is no need
for a stable superpartner.

The estimate of the baryon asymmetry produced by modular decay is simple.  Assume the amount
of baryon number produced per decay is $A$.  $A$ is the
product of a loop factor (presumably of order
$\alpha_s \over \pi$ times $CP$-violating
phases, which, given our assumptions, are
of order $1$).  The number of massless particles
produced per decay is ${m_M \over T_R}$ where $m_M$ is the mass of the modulus.
Plugging in the expected form of the modular mass and reheat temperature, we find a baryon 
to entropy ratio of
\eqn\basymm{{n_B \over n_{\gamma}} \sim A ({m_p \over f})^{\ha} {10^{11}\ GeV \over f}}
For an $f$ of order $m_p$ this gives a result in the desired range
if $A \sim 10^{-2} - 10^{-3}$.

For $f \sim 10^{16}$ GeV,
the baryon to entropy ratio is of order $ A 10^{- 4}$ 
so this
mechanism seems to produce too many baryons.  
Of course, this is the regime in which we might expect some
effect of electroweak baryon number violation, and the 
situation becomes much more complicated.
It seems clear that the simplest model
has $f \sim m_p$, a reheat temperature just above nucleosynthesis, 
baryogenesis from modular decay, and axions with decay 
constant $10^{16}$ GeV as dark matter.

We next explore the mechanism of coherent baryon number production of
ref. \affdine.  One can contemplate two possibilities
here.  The first is that among the moduli are the inflatons.  In that case,
we can take over directly the  estimates of the baryon
asymmetry from ref. \drt.  In that paper,  a formula
was given for the baryon number per inflaton.  Schematically,
this can be written:
\eqn\bperinflaton{{\rho_{\phi} \over \rho_I}\approx
({m_{3/2} \over m_p})^{2/(n-2)}}
where the term in the superpotential which lifts the
flat direction is of the form $\phi^n$, and we have
assumed that $m_p$ is the only scale.
To obtain the baryon to photon ratio
after reheating, one needs to multiply  this result by
$T_R/M_I$, where $T_R$ is the reheating temperature and
$M_I$ is the mass of the inflaton.  Assuming an inflaton
mass of order the weak scale and a reheat temperature
of order a few $MeV$, this factor is about $10^{-5}$.
So one sees that the flat direction must be extremely flat;
$n$ must be at least $8$.
As explained in ref. \drt, one can obtain directions this
flat -- or even exactly flat -- by means of discrete symmetries.

Alternatively, it might be that the inflaton reheat temperature is much
larger, and that moduli dominate the energy density for some
time.  In this case, the inflaton may  decay long before
nucleosynthesis.  However, unless the inflaton reheat temperature
is well above $10^{11}$ GeV, the moduli more or less
immediately come to dominate the energy density of the 
universe, and the estmate goes through  as above.
If the reheat temperature is higher (see the discussion
of the next section about the gravitino problem)
then if $R_B= {n_B/n_0}$, where $n_0$ is the entropy density
just after inflation, then
the final baryon to photon ratio is
\eqn\moduliplusinflaton{n_B/n_{\gamma}=R_B({T_R \over M_I})({T_m\over M_m}).}
Here $T_m$ denotes the reheat temperature after the
moduli decay, and $M_m$ denotes the moduli mass.
So in this situation, it will be necessary that the
potential in the flat direction be extremely flat, in order for coherent
baryon number production to be viable.

There are other possible mechanisms for generating the
baryon asymmetry which have been discussed in the literature,
such as $B$-violating gravitino decays\ref\vitures{J. Cline
and S. Raby, Phys. Rev. {\bf D43} (1991) 1781.}  In cases
where there is a stable LSP, one also needs to examine
LSP production\randallthomas.
These mechanisms could also be operative
here.  Many of the issues are reviewed in ref. \thomasbaryons.

\subsec{The Gravitino Problem}

All SUSY models have a potential problem with gravitino production in the early
universe.  Like moduli, gravitinos can dominate the energy density of the universe
and ruin the predictions of nucleosynthesis.  In inflationary cosmologies, gravitinos
are produced in the reheating of the universe through the decay of a coherent scalar
field.  In conventional inflationary scenarios, 
reheat temperatures greater than $10^9$ GeV or so lead to excessive production
of gravitini\ref\ellisetal{J. Ellis, J.E. Kim and
D.V. Nanopoulos, Phys. Lett. {\bf 145B} (1984) 181.}.  

The reheat of the universe through modular decay which we have discussed above,
is safely below this bound.  However, we must also worry about the possibility of
episodes of reheating that occurred prior to the period of cosmic history when
axions began to oscillate.  In particular, if we want to embed our scenario in
a model of inflation, we must enquire about the reheating due to the decay of
the inflaton.  

One possibility is that the modulus we have already discussed is itself the inflaton.
An apparent problem with this idea is that the modular energy density appears
to be much smaller that conventionally required for the inflationary explanation of
the magnitude of fluctuations in the cosmic microwave background.
There exist models
of inflation \ref\hybrid{A.D.Linde,
{\it Phys. Lett.}{\bf 259B}, 38 (1991); A.R.Liddle, D.H. Lyth, 
{\it Phys. Rep.} {\bf 231}, 1 (1993); A.D.Linde, {\it Phys. Rev.} {\bf D49}, 748, (1994); E.J.Copeland, A.R.Liddle,
D.H.Lyth, E.D.Stewart, D.Wands, {\it Phys. Rev.} {\bf D49}, 6410, (1994); E.D.Stewart, {\it Phys. Lett.} 
{\bf 345B}, 414, (1995).
} in which inflation at a scale of $10^{11}$ GeV can lead to
microwave fluctuations of the right size.  The most attractive of these models
\ref\supernatural{L. Randall, M. Soljacic,
and A.H. Guth, ``Supernatural
Inflation,"
MIT-CTP-2499, hep-ph/9601296.} , actually
seems to arise quite naturally in string theory at
intersection points of moduli spaces.  
Another way to reconcile the light modulus with inflationary expectations is to
assume that the modulus is a dilaton like field, on which the potential depends in
an exponential fashion (in a parametrization in which the Kahler potential of the
modulus is only slowly varying) in some extreme region of moduli space.  
Then one can imagine that inflation takes place in a region where the potential
is slowly varying, but the minimum is in the extreme region.  This explains the
discrepancy in inflation and SUSY breaking scales \ref\bbs{T. Banks,
M. Berkooz and P.J. Steinhardt,
Phys. Rev. {\bf D52} (19995) 705, hep-th/9501053.}.  One must confront the 
Brustein-Steinhardt problem\ref\brust{R. Brustein
and P.J. Steinhardt, Phys. Lett. {\bf B302}
(1993) 196.} in such a scenario, but as explained in the
appendix of \bbs\ this may not be too serious.

Suppose now that the inflaton is not the
modulus responsible for the Hot Big Bang.
It will have a large energy density,
and Planck scale couplings, giving it a reheat
temperature $T_I$ much higher than the scale of the
electroweak phase transition.
Reheating will produce a gravitino number density $n_G$ which is initially
of order
$n_g/s \approx 10^{-4} T_I/m_p$.
After inflaton reheating, the universe remains radiation dominated
for a while.  The gravitino energy density falls
like the cube of the scale factor, 
but grows linearly in the inverse temperature
relative to the radiation density.  The modular energy density remains constant
until the Hubble parameter is equal to the mass of the modulus.  If we choose 
a decay constant $f\sim m_p$, this occurs at a time when the total energy density
of the universe, and the modular density, are approximately equal.  The
corresponding temperature, $T_m$, is roughly
$T_m \approx 1/5 \sqrt {m m_p}$.
$n_g/s$ remains constant until the moduli finally decay.
With the assumption that the reheat temperature is
of order a few $MeV$, the gravitinos are diluted by
a factor of order $10^{14}$.
The gravitino to radiation density ratio grows by another
factor of $100$ before the gravitini decay.
Previous studies have shown that this ratio must be less than about
$10^{-7}$ in order for
gravitino decay to preserve the products of nucleosynthesis.  Thus
even for $T_I$ of order $m_p$,
the gravitino density is not a problem.   In other words
no matter what our assumptions about the nature of inflation,
the scenario outlined
in this section solves the gravitino problem.

\newsec{Small Violation of CP}

In string theory, CP is an exact (gauge) symmetry, so CP violation
is inevitably spontaneous.  There are numerous CP odd fields which
are candidates for breaking CP.
Our fundamental assumption in the present section, will be that the breaking of CP 
is small.  In order
to be precise we will work within the context of a specific model for the origin of
CP violation, but we believe that our results can be  generalized to other models
in a straightforward way.  The model that we
will use was proposed by Nir and Ratazzi \nir.  All CP violation can be traced
back to the vacuum expectation value (VEV) of a single field $S_3$, with magnitude
$<S_3> \sim \lambda^5 M_p \sim 3 \times 10^{-4} M_p \equiv \delta M_p$,
where $\lambda$ is the Cabibbo angle.
$S_3$ is a singlet under continuous
gauge groups, and has only nonrenormalizable couplings to fields with masses below
$M_p$.  The pattern of these couplings is determined by an abelian horizontal
symmetry.  In \nir\ it is shown that such a model can account for all CP violating
phenomena observed (and not observed) in nature.  In particular, it leads to a CKM
phase of order one, and constrains supersymmetric contributions to 
CP violating phenomena to be smaller than experimental upper limits.  It does not
by itself solve the strong CP problem.

Now let us turn our attention to the very early universe.  We make the standard
inflationary assumption that at early times the energy density of 
the patch which becomes our universe is dominated by nearly homogeneous classical
scalar fields.  We will adopt string theory language and call the space
of scalars the moduli space.  
Among these fields are the {\it inflatons}, which coordinatize
a submanifold of the moduli space to which the system's trajectory is rapidly
attracted, and on which it stays for many e-foldings of the universe.  The energy
density on this submanifold is approximately constant, and drives a quasi exponential 
expansion.  We will assume that this inflaton submanifold is approximately invariant
under CP, with the violation of CP being no larger than that in the vacuum.
Clearly this assumption is not necessary to an understanding of the small size of
CP violation in current experiments.  Its plausibility can only be judged within
the context of a theory of the potential on the space of fields, a theory which
does not yet exist.  

In general, during an inflationary era, we may expect the inflationary energy
density to depend on all of the scalar fields in the theory.  This is particularly
true in the context of supergravity \ref\dinetal{G. Dvali,
Phys. Lett. {\bf 355B} (1995) 78; M. Dine,
L. Randall, S. Thomas, ``Supersymmetry Breaking
in the Early Universe," Phys.
Rev. Lett. {\bf B75} (1995) 398, hep-ph/9503303.
Earlier references include
M. Dine, W. Fischler and D. Nemeschansky,
Phys. Lett. {\bf 136B} (1984) 179 and
O. Bertolami and G.G. Ross, Phys. Lett. {\bf 183B}
(1987) 163.}.  However, this argument
must be reexamined in the case of axions.  The very existence of the axion
depends on an approximate global symmetry of the theory, or at least of the
portion of moduli space in which the vacuum state resides.
The latter point of view is the one indicated by string theory.  Axion PQ
symmetries are approximate symmetries which
arise in string theory only in special regions of moduli space
where effective gauge couplings are small or internal dimensions large.
It is entirely plausible that these symmetries are much more
strongly broken on the inflaton
submanifold.  For example, the couplings of standard model gauge fields might
be very large on the inflationary submanifold.
In the context of
the models of \mtheory\ another way to break the axion symmetries in the early
universe is to shrink the eleventh dimension down to the size of the other
six compact dimensions.
Note that the assumption of PQ symmetry
breaking during inflation contrasts with our assumption of approximate CP
symmetry. 
Needless to say, determining what actually happens requires a much
better understanding of moduli dynamics than we currently
possess.

How large might the axion potentials be?  Consider, first, axions
whose potential vanishes as $R \rightarrow \infty$.  We can
easily imagine that $R = {\cal O}(1)$ during inflation.
In this case, the axion mass would be expected
to be of order the Hubble constant.   For example, one
might expect terms in the superpotential of the form
\eqn\axionestimatea{ W= e^{-R} I}
where $I$ is the inflaton field (assumed to have a non-vanishing
$F$-component).  This leads to
\eqn\axionestimateb{m_a^2 \approx H e^{-R}/f_a^2.}
For the field which at weak coupling is termed the
``model-independent axion," and usually denoted $S$,
one expects a similar result.  This corresponds to the possibility
that the QCD coupling ($\alpha_s$) is of order one during
inflation.

We note that there is another puzzle of supersymmetric inflationary cosmology which
may be resolved by the assumption that couplings and scales were all of order one
(in fundamental units) during inflation.  Typical inflation models require
the inflationary energy density to be much larger than the square of the
SUSY breaking order parameter in the vacuum.  If the low scale of SUSY
breaking is explained by a small coupling, then this discrepancy is
removed \bbs .  We have already invoked this mechanism in the previous section.

Given these assumptions, the evolution of the axion during inflation can
be described simply.  Let $a = 0$ be the CP invariant value of the axion field.
During inflation, the axion feels an effective potential which gives it an
effective mass of order the Hubble constant.  It is rapidly driven to
a minimum of this potential, which lies at a distance ${a\over f_a} \sim \delta$ from
the origin.  By the approximate CP symmetry,
this is the same order of magnitude as the distance to the true minimum
of the vacuum axion potential.

The postinflationary evolution of the universe depends on whether or not there
are light moduli of the sort we described in the previous section.  The strongest upper bound comes from assuming that there are
no such moduli.  As usual we assume that
the reheat temperature of the inflaton is well above the QCD scale
and the postinflationary universe contains only axions and radiation.
This is the conventional scenario for axions, with the exception of the fact
that the initial distance of the axion field from its minimum is one ten thousandth of
that which is usually assumed.  For the axion energy
density one has\axioncosmologya\axioncosmologyb\axioncosmologyc:
\eqn\axionenergy{\Omega_a h^2 \approx 0.7 \theta_o^2
\times (f/ 10^{12}\rm GeV)^{1.18}.}
(There are uncertainties in these formulas of perhaps an
order of magnitude.)
If $\theta_o$, the initial value of $\theta$,
is of order $10^{-3}$-$10^{-4}$, then we can tolerate
decay constants as large as $m_p$. This is the order of
magnitude of expected for $\theta_o$ in models of small
CP violation, in which $\theta_o \sim \delta$.

If some portion of early cosmological history is dominated by coherent scalars with
masses coming from susy breaking then we can repeat the analysis
of the previous section, but with 
initial axion energy density smaller by a factor of $\delta^2 \sim
10^{-7}$ .  Eqn. \bound\ now becomes
\eqn\boundnew{\delta^2{f_a^2 \over m_p^2}{T_R \over {10\ eV}} \leq 1}
For a given modular reheat temperature the upper bound on the 
axion decay constant is larger by a factor $\delta^{-1}$. For
$\delta \sim 10^{-4}$, this again gives a limit of order $m_p$.
In each case, one can choose parameters so that
axions constitute the dark matter in the universe.   

\newsec{\bf Eleven Dimensional Physics}

Our original motivation for returning to the cosmological conundra of axions was the
observation that the fit of strongly coupled heterotic string theory to available
data\ref\witten{E. Witten, ``Strong Coupling Expansion of
Calabi-Yau Compactification, hep-th/9602070.} 
leads to a general prediction of QCD axions with large decay
constants\ref\banksdine{T.Banks, M.Dine, ``Couplings and Scales in
Strongly Coupled Heterotic String Theory", hep-th/9605136.}.
We would like to devote this section to a brief discussion of
the cosmology of these strongly coupled string vacua.

Four dimensional vacua of the strongly coupled $E_8$ X $E_8$ 
heterotic string with $N=1$ SUSY, are 
described by M-theory (for practical purposes eleven dimensional supergravity) 
compactified on a seven manifold with boundary which is a Calabi-Yau fibration over 
an interval.  The fit \witten to the unified fine structure constant, Newton's
constant, and the unification scale, in terms of the fundamental eleven dimensional
Planck scale $l_p$, and the geometry of the manifold is as follows.  On the boundary of
the manifold with standard model gauge group, the Calabi-Yau size $R$ (volume $= R^6$), 
which determines the unification scale, 
is twice $l_p$, and the length of the interval,$R_{11}$, is about $70 l_p$.  The Calabi-Yau
volume decreases monotonically along the interval, and as a consequence the bare
coupling of the hidden $E_8$ is much larger than that of the standard model.
In this regime, the heterotic string is more properly thought of as a membrane
stretched between the two boundaries.  Its tension is $\sim 70 l_p^{-2}$.

Axions are modes of the three form gauge field of M theory which have the form $a(x) 
b_{MN}(x^{11}) dx^M dx^N dx^{11}$, where $a$ depends only on the noncompact coordinates,
and $b_{MN}$ is one of the harmonic $(1,1)$ forms on the Calabi-Yau manifold at $x^{11}$.
These are almost pure gauge modes when $a$ is a constant.  They can be written as
$dC$, where $C$ is a two form which vanishes on the $E_8$ boundary.  Thus, constant
shifts of $a$ are Peccei-Quinn symmetries broken only by effects on the standard model
boundary.  As argued in \banksdine , the strongest such effect is QCD.

In the limit $R_{11} \gg l_p$, which seems like a reasonably good approximation to 
the real world, the kinetic terms of the axions are given by their values in the
Kaluza-Klein approximation to M theory dynamics.  As noted in  \banksdine, the gauge coupling functions
and Kahler potentials of the bulk moduli are the same
in the weak and strong coupling limits, so the
result is identical
to that  of Kim and Choi. 
Thus, the axion of strongly coupled heterotic string theory fits fairly well into the
framework of our first scenario.  We note \banksdine\ that the true QCD axion in the strongly
coupled region may in fact originate as a gauge bundle modulus on the standard model
boundary.  In this case, the order of magnitude estimate of the axion decay constant
is $l_{11}^{-1} \sim 2 \times 10^{16}$ GeV.  We have not yet been able to understand the precise
numerical factors in this formula.  If the two decay constants have the same order of 
magnitude the QCD axion will be a linear combination of the mode coming from the three
form, and the boundary modulus.

The other part of our scenario is a modulus with a decay constant of order $m_p$ and
potential energy of order the intermediate scale.  Paradoxically, although $m_p$ is the 
natural order of magnitude estimate of the decay constant of a bulk modulus we can be
less than sure that such a modulus exists.  The axionlike moduli seem to have much smaller
decay constants.  The Kahler potentials of other moduli are not easily calculable
for the values of the moduli which seem to fit the data.  Our scenario looks very natural
in the eleven dimensional context, but we cannot calculate enough parameters to be sure
that it is realized.

Another general issue which we must face is the question of whether a satisfactory model
of inflation can be built in this region of moduli space.  The danger is that a
simpleminded estimate of density fluctuations indicates that one needs a vacuum energy
of order $M^4$ with $M \sim 10^{16}$ GeV in order to account for the COBE data.  
This is uncomfortably close to the eleven dimensional Planck scale, above which we
have no description of the correct physics in the M theory regime.  In fact,
satisfactory models of inflation with  vacuum energy scale lower than $M$ by an order
of magnitude abound in the literature.  Theoretically, the order of magnitude estimate
of the maximum potential energy of bulk moduli in the M theory regime is (assuming
modular decay constants of order $m_p$)
$l_{11}^{-6}m_p^{-2} $,  $10^{-4}$ times smaller than the eleven dimensional
Planck density which marks the border of our ignorance. This gives an effective $M$ of
order $2 \times 10^{15}$ GeV.  It is conceivable then that a semiclassical model of
inflation which is both theoretically and phenomenologically viable, can be built
in this regime.

The results of \banksdine\ also tell us something about the question of whether the
moduli whose decay gives rise to the Hot Big Bang are inflaton fields.  In the
strongly coupled heterotic string
it proved difficult to generate a nonperturbative energy scale higher than the hidden
sector scale which gives rise to SUSY breaking.  One had to assume that the
the vacuum value of the modulus ${\cal S}$ which controls the
hidden sector coupling, was at small values of the coupling.  A natural way\bbs
to explain the large vacuum energy during inflation is to assume that ${\cal S}$
is one of the inflatons and that inflation occurs at strong coupling where the
estimate of the energy density given in the previous paragraph is valid.  
On the other hand, the mass and apparent couplings of ${\cal S}$ make it a
candidate to be the progenitor of the Hot Big Bang.  Thus, the eleven dimensional
scenario is likely to identify the two fields, and consequently has a very low
reheat temperature.

This has been a mere sketch of the cosmology of strongly coupled heterotic string
theory.  We hope to return to the subject when more is understood about the effective
lagrangian of the important scalar fields in this region of moduli space.

\newsec{Conclusions}

We have exhibited two cosmological scenarios which significantly modify the bound 
on the axion decay constant.  In combination with the results of \choi , which
expand the range of expected values of $f_a$ in superstring theory, it is now possible
to claim that superstring axions may be compatible with all experimental and 
cosmological data.  Our first scenario also points up the fact that the cosmological
moduli problem may be much less significant than we had imagined.  Moduli compatible
with weak scale baryogenesis are at the limits of plausibility but nucleosynthesis
is certainly not a problem.  Alternate schemes for baryogenesis are available and lead
to an attractive and consistent picture of the cosmology of the very early universe.
It is likely that axions with decay constant $10^{16}$ GeV will be the dark matter
in such a scenario.

Our second solution of the cosmological axion problem has less to say about superstrings
and moduli, but may have more immediate implications for low energy physics.

Our analysis of the first scenario relied on the hidden sector mechanism for SUSY breaking.
It is worth saying a few words about axion properties in models where SUSY is
dynamically broken in sectors coupled to the standard model through gauge interactions.
In such a model, all moduli which get their mass from SUSY breaking are extremely light and
have lifetimes longer than the current age of the universe.  There are no apparent 
candidates for fields whose decay would initiate the Hot Big Bang at temperatures of
order the weak scale or below.   Thus, the first scenario is completely untenable in
such a model.  Worse, all of the moduli now tend to dominate the energy density of
the universe from an energy density ${f^2 \over m_p^2} F^2$ (here $f$ is a typical
modular decay constant, and $F$ is the SUSY breaking order parameter), until long after
the present era.  One is forced to imagine that the true vacuum sits at a point invariant
under symmetries that transform all of the light moduli.  In such a case, one might
invoke a version of our second scenario to force all of the moduli to their true minima
during inflation.  It is however much less plausible that the inflaton is invariant
under the large group of symmetries required to fix all of the light moduli.  If it is not,
then the scenario does not work.  

Finally we note that scenarios in which the very early universe was dominated by coherent
scalar fields eliminate many hypothetical cosmological phase transitions.  In particular,
domain walls due to spontaneously broken discrete symmetries might be a problem only
if they are produced in the brief period between inflaton reheat and modular dominance
(in scenarios in which the inflaton is not the modulus which produces the hot big bang), 
or after modular reheat.

\vskip 1cm
{\bf Acknowledgments:}
We thank K. Choi for discussions and for
calling our attention to ref. \choi.
We also thank S. Thomas for important
critical comments.
The work of T. Banks was supported in
part by the Department of Energy under
grant $\# DE-FG0296ER40559$.
The work of M. Dine is supported
in part by the U.S. Department of Energy.

\listrefs
\bye
\end